\def\spose#1{\hbox to 0pt{#1\hss}}
\def\approxlt{\mathrel{\spose{\lower 3pt\hbox{$\sim$}}
        \raise 2.0pt\hbox{$<$}}}
\def\approxgt{\mathrel{\spose{\lower 3pt\hbox{$\sim$}}
        \raise 2.0pt\hbox{$>$}}}
\def\approxpropto{\mathrel{\spose{\lower 3pt\hbox{$\sim$}}
        \raise 2.0pt\hbox{$\propto$}}}
\mathchardef\twiddle="2218

\def\multleft#1{\hbox to size{\vbox {\halign {\lft{##}\cr #1}}\hfill}\par}
\def\multright#1{\hbox to size{\vbox {\halign {\rt{##}\cr #1}}\hfill}\par}
\def\Mdot{\hbox{$\dot M$}}
\def\mdot{\hbox{$\dot m$}}
\def\<{\thinspace}

\def\apc{\rm atom cm$^{-2}$}

\def\erg{{\rm\thinspace erg}}

\def\g{{\rm\thinspace g}}

\def\keV{{\rm\thinspace keV}}

\def\kpc{{\rm\thinspace kpc}}

\def\s{{\rm\thinspace s}}

\def\ergps{\hbox{$\erg\s^{-1}\,$}}

\def\apc{\rm atom cm$^{-2}$}

\newcommand\beq{\begin{equation}}
\newcommand\eeq{\end{equation}}

\documentstyle[11pt,aaspp4,psfig]{article}

\begin{document}
\title{Accretion onto the Supermassive Black Hole in M87}
\author{Tiziana Di Matteo\altaffilmark{1,2,3}, Steven W. Allen\altaffilmark{4},
Andrew C. Fabian\altaffilmark{4},  Andrew S.Wilson\altaffilmark{5},  
Andrew J. Young\altaffilmark{5}}
\altaffiltext{1}{Harvard-Smithsonian Center for Astrophysics, 60
 Garden St., Cambridge, MA 02138} 
\altaffiltext{2} {Max-Planck-Institute f{\" u}r Astrophysik, Karl-Schwarzschild-Str. 1, 85740 Garching 
bei M{\" u}nchen, Germany}
\altaffiltext{3}{Carnegie-Mellon University, 
Dept. of Physics, 5000 Forbes Ave., Pittsburgh 15231}
\altaffiltext{4} {Institute of Astronomy, Madingley Road, 
Cambridge, CB3 OHA, UK }
\altaffiltext{5} {Astronomy Department,
 University of Maryland, College Park, MD 20742 }

\begin{abstract}
{\it Chandra} X-ray observations of the giant elliptical galaxy M87
resolve the thermal state of the hot interstellar medium into the
accretion (Bondi) radius of its central $3 \times 10^9 M_{\odot} $
black hole. We measure the X-ray gas temperature and density profiles
and calculate the Bondi accretion rate, $\dot{M}_{\rm Bondi} \sim 0.1
M_{\odot}$yr$^{-1}$. The X-ray luminosity of the active nucleus of M87
observed with {\it Chandra} is $L_{x, 0.5-7 \keV} \sim 7 \times
10^{40}$ erg s$^{-1}$. This value is much less than the predicted
nuclear luminosity, $L_{\rm Bondi} \sim 5 \times 10^{44}$ erg
s$^{-1}$, for accretion at the Bondi rate with a canonical accretion
radiative efficiency of 10\%. If the black hole in M87 accretes at
this rate it must do so at a much lower radiative efficiency than the
canonical value. The multiwavelength spectrum of the nucleus is
consistent with that predicted by an advection-dominated flow.
However, as is likely, the X-ray nucleus is dominated by jet emission
then the properties of flow must be modified, possibly by outflows.
We show that the overall energetics of the system are just consistent
with the predicted Bondi nuclear power. This suggests that either most
of the accretion energy is released in the relativistic jet or that
the central engine of M87 undergoes on-off activity cycles.  We show
that, at present, the energy dumped into the ISM by the jet may reduce
the accretion rate onto the black hole by a factor $\propto
(v_j/c_s)^{-2}$, where $v_j$ is the jet velocity and $c_s$ the ISM
sound speed, and that this is sufficient to account for the low
nuclear luminosity.
\end{abstract}

\keywords{accretion, accretion disks --- black hole
physics --- galaxies:individual (M87) --- X-rays: galaxies}

\section{Introduction}

The nucleus of M87 (NGC~4486) contains a black hole of mass $ M \sim 3
\times 10^9 M_{\odot}$ directly determined from {\it HST} observations
(Ford et al. 1995; Harms et al. 1995; Macchetto et al. 1997). It is
a nearby active nucleus with a one sided jet and large-scale radio
structure. However, the activity displayed by its nucleus is far less than
what is predicted if the central black hole were accreting mass from its
hot interstellar medium (with a standard radiative efficiency of $\sim
10$ per cent; see e.g. Fabian \& Rees 1995; Reynolds et al. 1996; Di
Matteo et al. 2000). M87 is probably the most illustrative case of a
low-luminosity system otherwise common in nearby galaxies known to
contain supermassive black holes (e.g, Magorrian et al. 1998;
Ferrarese \& Merritt 2000).

There are two possible explanations for the low luminosities of nearby
black holes: (a) the accretion occurs at extremely low rates or (b)
the accretion occurs at low radiative efficiencies as predicted, for
example, by advection dominated accretion flow models (ADAFs;
e.g. Rees et al. 1982; Narayan \& Yi 1994,1995; Abramowicz et
al.~1995 but see also inflow-outflow models; e.g. Blandford \&
Begelmann 1999; Stone et al. 1999; or CDAFs; e.g. Quataert \& Gruzinov
2002). In order to discriminate between these two possibilities,
it is necessary to determine both the accretion rates and the nuclear
luminosities precisely. Direct measurement of the latter and
evaluation of the former has not been possible with previous X-ray
satellites. Here, we show that, for the nucleus of M87, both can now
be achieved with {\it Chandra}.

The massive black holes at the center of elliptical galaxies are
likely to accrete primarily from the surrounding hot, quasi-spherical
ISM. Accretion rates can therefore be simply estimated using Bondi
accretion theory. This requires accurate measurements of both the
density and temperature of the hot, X-ray emitting ISM at the 'Bondi
accretion radius', the radius at which the gravitational potential of the
central black hole begins to dominate the dynamics of the hot gas and
the gas starts to fall into the black hole.

Thanks to its high spatial resolution and sensitivity, the {\it
Chandra X-ray Observatory} is able to provide some of the most
stringent constraints on the properties of low-luminosity black
holes. In particular, at the distance of M87 (18 Mpc), the spatial
resolution of {\it Chandra} corresponds to a radius of less than a
hundred parsec or, equivalently, a few $10^{5}$ Schwarzschild
radii. For M87, this allows us to measure, for the first time,
fundamental properties of the ISM {\it at the accretion radius} of the
black hole, and thereby estimate the mass supply into the
accretion flow.

\section{Gas properties at the black hole accretion radius}
\subsection{Central temperature and density}

Details of the Chandra observations and observing strategy are
presented by Wilson \& Yang (2002). In brief, M87 was observed twice
with {\it Chandra}, on 2000 July 29 and 2000 July 30. Both
observations  were carried out using the Advanced  CCD Imaging
Spectrometer (ACIS) and back-illuminated S3 detector. The first
observation was made in a full-frame mode with a 3.2 frame time. The
net exposure time, after removing periods associated with small
background flares, was 33.7ks. These data have been used to study the
properties of the diffuse X-ray gas in M87. The second, shorter
observation was made in a reduced 1/8 sub-array mode with a frame time
of 0.4s. The use of this mode  avoids problems associated with pile-up
in the data for the brightest knots of the jet and the central AGN
(Wilson \& Yang 2002). The exposure time for the second observation
was 12.8ks. High resolution {\it Chandra} images and detailed
analyses of the jet and cluster properties are presented by Wilson \&
Yang (2002) and Young, Wilson \& Mundell (2002), respectively.

Spectra were extracted from five circular annuli centered on the
nucleus (Figure 1). The data for the innermost region, which has a
radius of 4 detector pixels (approximately 2 arcsec or 0.2 kpc at the
distance of the source), were extracted from the shorter 12.8 ks data
set. The emission from this region is dominated by the central
AGN. The data for the four outer annuli (covering radii of 0.2-1, 1-4,
4-8 and 8-12 kpc, respectively) were extracted from the longer,
full-frame observation. The spectra were grouped to contain a minimum
of 20 counts per pulse invariant channel, allowing $\chi^2$ statistics
to be used. Background spectra were extracted from the blank-field
data sets provided by the Chandra X-ray Center. Separate
photon-weighted response matrices and effective area files were
constructed for each region using the appropriate ACIS-S calibration 
and response files. We have used the ACISABS model in XSPEC to 
account for the time varying molecular contamination of the ACIS optical 
blocking filters.

The spectra have been analyzed using the XSPEC code (version 11.2.01;
Arnaud 1996) and following the deprojection method described by Allen
et al. (2001). We assume that the diffuse X-ray gas in each of the
spherical shells, defined by the annuli mentioned above,
is isothermal and can be described by a vmekal model (Kaastra
\& Mewe 1993; incorporating the Fe L calculations of Liedhal,
Osterheld \& Goldstein 1995) with the abundances of O, Mg, Si, S and
Fe included as independent free fit parameters. The fit to the
outermost annulus is used to determine the temperature, emission
measure and abundances of the ISM in the outermost spherical
shell. The contribution from that shell to each inner annulus is then
determined by purely geometric factors. The fit to the second annulus
is used to determine the parameters for the second spherical shell,
and so forth, working inwards. For the central 2 arcsec region,  we
have included an additional power-law component to model the emission
from the AGN (see Section 3). We assume that the intrinsic spectrum of the
ISM in this region is the same as in the next annulus out.
The emission from all annuli is assumed to be absorbed by a column
density, $N_{\rm H}$, of cold gas, which is a single, additional free
parameter in the fits.  The emission from the brightest knots of the
jet and all other point sources were masked out and excluded from the
analysis.

The deprojected temperature profile of the X-ray gas, determined from
a simultaneous fit to the five annular regions, is shown by the solid
points in Figure 2. The deprojected temperature decreases from
$kT=1.83\pm0.01$ keV at $r= 10$ kpc to $kT=0.80\pm0.01$ keV within the
central kpc ($1\sigma$ errors). The total $\chi^2$ for the deprojected
fit of 3171 for 1271 degrees of freedom indicates that the simple,
spherically-symmetric model provides an incomplete description of the
Chandra data, which is unsurprising given the structural complexities
evident in Fig. 1 (see also Wilson \& Yang 2002 and Young et
al. 2002). Nevertheless, the temperature measurements should provide a
reasonable description of the mean, emission-weighted properties in
each spherical shell (although the uncertainties may be underestimated
by a factor $\sim 2$). We note that the temperature measurement for
the $0.2-1$ kpc annulus is not affected significantly by flux from the
central AGN, scattered outwards in the wings of the point spread
function. In order to test this we carried out simulations in
XSPEC, using our deprojected model, in which we added a power-law
component to the model for the first annulus out (0.2-1kpc), with a strength
calculated according to the observed nuclear flux and PSF models in
the Chandra handbook. We find that the flux in the wings of the PSF
has no significant effect on the best-fit parameters for the 0.2-1kpc
annulus. To be conservative, we then doubled the strength in the wings
and still found no significant change in the best-fit parameters, 
indicating that our results are robust against this effect.

Figure 2 also shows the importance of deprojecting the X-ray data when
determining the central gas temperature (and hence the density
profile). Without deprojection, we would infer a spuriously high
temperature of $kT=1.16\pm0.01$ keV for the $0.2-1.0$ kpc region,
which would impinge on the calculations of the accretion rate in
Section 2.2.  Note that deprojecting also improves the total $\chi^2$
for the fit by 410 compared with a direct analysis of the projected
spectra, without including additional free parameters in the fit.

The electron density profile for the X-ray gas, shown in Figure 2, was
derived by deprojecting the surface brightness profile, given the
deprojected temperature profile. At radii larger than a few kpc the
density profile follows $\rho \approxpropto r^{-1}$.  Within the
central $\sim 2$ kpc, the density profile flattens: fitting a constant value
to the central two annuli we obtain $n_e=0.170 \pm 0.003$ cm$^{-3}$
(or $n_e = 0.163 \pm 0.007$ cm$^{-3}$ for the innermost annulus).

\subsection{$\dot{M}$}
We can estimate the accretion rate from the ISM onto the central black
hole using Bondi accretion theory; the nuclear accretion rate
will thus be determined by the density and temperature of the hot gas at
the point where the influence of the black hole becomes dominant.  In
the standard adiabatic problem (see Appendix A with $H=0$ for the
exact Bondi results), the Bondi (accretion) solution implies the
sphere of influence of the black hole extends out to the
gravitational capture radius (see also e.g.; Di Matteo et al. 1999;
Quataert \& Narayan 2000; Nulsen \& Fabian 2000):
\begin{equation}
r_A \sim b \frac{GM}{c_s^2} = 
0.05 T^{-1}_{0.8} M_9 \; \hbox{\rm kpc} \simeq 5 \times 10^5 r_{\rm Sch}. 
\end{equation}
Here $M_9 = 10^9 M_{\odot}$ is the mass of the black hole ($M_{9} =
3$), $T_{0.8} = T/0.8 \;\keV$ is the temperature in \keV (cf. Figure
2), $c$ is the speed of light, $c_{\rm s} \sim 2.7
\times 10^7T_{0.8}^{1/2}$ cm s$^{-1}$ is the sound speed (where $T$ is
the ISM gas temperature) and $r_{\rm Sch} = 2GM/c^2$ is the
Schwarzschild radius of the black hole. The parameter $b$ depends on
the detailed physics close to the accretion radius (including
$\gamma$). 
We initially take $b \sim 1$, for which our definition of $r_{A}$
corresponds to $\sim 2 r_{s}$, the sonic radius of the black hole, and
discuss this further in \S 5.1 and in the Appendix. 
Note that, depending on the definition one chooses,
$r_{A}$ lies approximately at $ 0.5 - 2 GM/c_s^2$. 
$r_{A} \approxgt 0.2 GM/c_s^2$ is resolved by {\it Chandra}.

Equation 1 shows that the accretion radius of M87 (for an assumed
distance of 18 Mpc) corresponds to an angular scale of 2 arcsec.
Thus, using {\it Chandra}, we are able to measure the properties of
the gas at approximately the black hole accretion radius for M87. We
note that the ionized disk of gas seen by HST (Harms et al. 1994) has
been measured from 0.25 arcsec to about 2 arcsec from the nucleus,
close to the Bondi radius of M87. HST, thus may have imaged the
accretion flow in the optical I-band.

The accretion rate is related to the density and temperature at the
accretion radius by the continuity equation:
\begin{equation} 
\dot{M}_{\rm Bondi} = 4\pi r_{\rm A}^2\rho_{\rm A}c_{\rm s}(r_{\rm A}),
\end{equation}
where $\rho_{\rm A}$ and $c_{\rm s}(r_{\rm A})$ are the density and
sound speed at the sonic radius.
With the density and temperature measured directly
at the accretion radius (Section 2.1), the Bondi accretion rate to M87
is given by
\begin{eqnarray} \dot{M}_{\rm Bondi}& = &7 \times 10^{23} \;
M_9^2\;\;T_{0.8}^{-3/2} \;n_{0.17} \;\;\;\; \hbox{$\rm\thinspace 
g s^{-1}$} \nonumber \\
&&\simeq 0.1\;\; \hbox{$\rm\thinspace M_{\odot} yr^{-1}$}. 
\end{eqnarray} 
where $n_{0.17} = n/0.17$ cm$^{-3}$.
Eqn.~(3) provides the accretion rate at the outer edge of the
accretion flow. As discussed in Section 5, the mass accretion
rate onto the black hole may be smaller if e.g.~it decreases with
radius because of an outflow (cf. Blandford \& Begelman 1999).

The Bondi accretion rate (Eq. 3) implies a luminosity
\begin{equation}
L_{\rm Bondi}=\eta \dot{M}_{\rm Bondi}
c^2 \simeq 5 \times 10^{44} \hbox{\rm erg s$^{-1}$}
\end{equation}
if $\eta =0.1$, as in a standard, radiatively efficient thin disk.

\section{The X-ray Luminosity}
The properties of the central AGN have been determined using the data
for the central 2 arcsec (4 pixels) radius region after accounting for
the emission from diffuse X-ray gas viewed in projection against the
nucleus (see \S 2.1). We find that the central point source can be
described by a power law model with a photon index $\Gamma =
2.23\pm0.04$ and a flux density at 1 keV of $(8.0 \pm 0.2) \times
10^{-13}$ erg cm$^{-2}$ s$^{-1}$ keV$^{-1}$ (1$\sigma$ errors). We
find no evidence for excess absorption associated with the central AGN
($\Delta N_{\rm H} < 3.2 \times 10^{20}$\apc~at 3$\sigma$ confidence),
over and above the mean value measured across the central 12 kpc
radius region. We note that the mean absorbing column density 
measured across the central 12 kpc of $N_{\rm H} = 1.97\pm0.12 \times
10^{20}$\apc~is slightly lower than the nominal Galactic value of $2.5 \times
10^{20}$\apc~determined from the HI studies of Dickey \& Lockman
(1990). This deficit may reflect residual systematic uncertainties in
the effective area of the instrument at low energies. Our results on
the central AGN are in averall agreement with those of Wilson \& Yang
(2002; although these authors find a higher absorbing column density
for the nucleus than the jet knots).  Our measured photon index is
slightly lower than the value of Marshall et al. (2001). (Note that
our analysis differs from that of Wilson \& Yang in that we account
for the emission from diffuse X-ray gas viewed in projection against
the nucleus.)

The observed nuclear luminosity is more than four orders of  magnitude
smaller than the predicted Bondi luminosity, implying (unless the
Bondi estimate is inappropriate) that the radiative efficiency $\eta
\sim 10^{-5}$. In the next section, we examine the predictions of hot
accretion flow models with low radiative efficiencies.

\section{Accretion models}

A hot accretion flow around a supermassive black hole will radiate in the
radio to X--ray bands. In the radio band, the emission results from
synchrotron radiation. At higher energies, and up to the X-ray band, the
emission is produced by bremsstrahlung processes and inverse Compton
scattering of the soft synchrotron photons (e.g.~Narayan, Barret \&
McClintock 1998).

In this Section, radii in the flow are written in Schwarzschild 
units: $r = ar_{\rm Sch}$.
We write black hole masses in solar units and accretion rates in
Eddington units: $M_{\rm BH} = m M_{\odot}$ and $\dot{M}= \dot{m}
\dot{M}_{\rm Edd}$. We take $\dot{M}_{\rm Edd} = 10L_{\rm Edd}/c^2 =
2.2 \times 10^{-8} m M_{\odot}$ yr$^{-1}$, i.e., with a canonical 10\%
efficiency. We take $a=10^4$ to be the outer radius of the flow. The
Bondi accretion rate for M87 in these units is $\dot{m}_{\rm Bondi} = 
1.6 \times
10^{-3}$.

The predicted spectrum from an ADAF depends (weakly) on the ratio of the
gas to magnetic pressure $\beta$, the viscosity parameter $\alpha$,
and the fraction of the turbulent energy in the plasma which heats the
electrons, $\delta$. Here, we fix $\alpha = 0.1$, $\beta = 10$, and
take $\delta = 0.3$ or $0.01$. The two major parameters, though, are
the accretion rate $\dot{M}$ and the black hole mass $M$, both of
which are constrained. With $M$ given for M87, we normalize the models
to the observed {\it Chandra} flux. This gives us the $\dot{m}$
required by the models to explain the X-ray emission. A model is
ruled out if it requires $\dot{m} << \dot{m}_{\rm Bondi}$ to
account for the observed luminosity or, equivalently, if we take 
$\dot{m} = \dot{m}_{\rm Bondi}$ and the models predict a higher
luminosity than is observed.

The solid dots in Figure 3 are the high resolution VLBI, HST, Gemini,
Keck (at 10$\mu m$) and Chandra high resolution measurements of the
nuclear flux in the radio, optical, mid-infrared and X-ray bands,
respectively. The open dots show the lower resolution VLA radio
measurements which are likely to include a more significant
contribution from the jet. The solid line shows the predicted spectrum
for a pure inflow ADAF model adjusted to roughly match the 1 keV {\it Chandra}
flux. In this model $\dot{m} = 6
\times 10^{-4}$, (with $\delta =0.3$) which is consistent with the
Bondi estimate (i.e. within the model uncertainties, which should be
taken to be $\sim 50 $\%). A model with $\delta =0.01$ and a
corresponding $\dot{m} = 10^{-3}$ is shown by the dotted line. The
required accretion rates are always large enough for Comptonization of
the synchrotron emission to dominate the X-ray emission in these
models (where the scattering optical depth in an ADAF is $\propto
\mdot$; see e.g., Narayan et al. 1998). The exact positions of the
Comptonization bumps in the optical and X-ray bands are a
function of temperature and, therefore, $\delta$. Models with higher values of
electron heating, which are also preferred from a theoretical point of
view, agree better with the measured 2--10 keV spectral slope
of $\Gamma \sim 2.2$. Note that, as shown in Figure 3, the agreement
with the observed spectrum is obtained only if the Compton bumps are
pronounced, i.e. the agreement depends upon 
the details of the model. 

Marshall et al. (2002) and Wilson \& Yang (2002) have shown that the
spectrum of the innermost jet knots observed by {\it Chandra} are
consistent with that of the nucleus, possibly implying a similar
origin. It is indeed plausible that a large fraction of the observed
nuclear X-ray flux might be due to emission from the base of the
jet\footnote{Note that it is not possible to put strong constraints on
the nuclear variability from these two observations because of the
different apertures considered in the published nuclear
measurements. However, there does not seem be evidence for substantial
variability. Evidence for such variability would argue against simple
ADAF models}. The point of Figure~3 is to show that ADAF models are
consistent with the requirement of a low-radiative efficiency. In
fact, Figure 3 shows that it is possible for the accretion flow to
account for a large fraction of the observed nuclear emission, thereby
offering an alternative explanation.

\section{Discussion}
{\it Chandra} observations of the nucleus of M87 have resolved the
properties of the ISM all the way into the Bondi radius of the central
black hole. This makes M87 the only extragalactic black hole system in
which both the gas temperature and density are measured into
approximately the accretion region, and for which a Bondi accretion
rate can be directly calculated (see Narayan 2002 or Quataert 2002 for
similar study of Sgr A$^*$). M87 also possesses a relatively bright
nuclear source, the flux of which can be accurately measured in the
X-ray band with {\it Chandra}, and throughout the whole of the
spectral energy distribution with other observations. At present, M87
is the best constrained low-luminosity extragalactic black hole
system. With both $L_{x}$ and $\dot{M}$ measured, we were able to show
unambiguously that the black hole in M87 is highly underluminous with
respect to its Bondi mass supply: the expected power output exceeds,
by about 4 orders of magnitude, the observed value.  If the black hole
in M87 is indeed accreting close to its Bondi rate (cf., \S 5.1), the
accretion efficiency, $\eta$, must be low. We have shown that the
required values of $\eta$ are consistent with predictions from ADAF
models.

We note that our general result is independent of whether the observed
nuclear luminosity is attributed to an ADAF or entirely dominated by
jet emission (see e.g. Wilson \& Yang 2002; Marshall et al. 2002).
The constraints derived here show that it is possible for a
significant fraction of the observed nuclear emission to be
contributed by the accretion flow. More importantly, perhaps, the
requirement for low-radiative efficiency of the accreting gas can only
be invalidated if the actual mass accreted by the black hole were to
be a very small fraction of the mass supplied in the outer region of
the flow (i.e. of $\dot{M}_{\rm Bondi}$). This could be the case if
strong outflows or perhaps convection are important in the accretion
flow, as has been discussed by e.g. Blandford \& Begelman (1999), Stone et
al. (1999) and Quataert \& Gruzinov (2000), Narayan et al. (2000),
respectively (but see Balbus \& Hawley's [2002] discussion on
convection). In such models, the accretion flow would produce a much
lower luminosity than a pure inflow ADAF and make a negligible
contribution to the observed luminosity (which, in the case of M87,
can be easily explained by the jet emission. In the CDAF case, the
energy would probably need to be tapped from the black hole's spin in order
to explain the high jet power). The Bondi rate could also be
decreased if the jet heats the ISM. We discuss this further in
\S 5.1.

In a previous study using radio observations, Di Matteo et al. (1999;
2000) favored outflow models for the nuclei of ellipticals.  With {\it
Chandra}, the properties of the central regions of the ISM gas have
been determined more accurately for a number of elliptical galaxies
(e.g., Loewenstein et al. 2001). In particular, the new observations
have shown that the ISM density profiles tend to flatten off
significantly in the central regions of ellipticals (as is the case in
M87 - Figure 2), implying Bondi accretion rates smaller than was
previously estimated. In most cases, {\it Chandra} does not detect
nuclear point sources (Loewenstein et al. 2001). Given that their
black hole masses are more uncertain, the constraints for
these other low luminosity systems remain inconclusive. The only two
{\it Chandra} detections of X-ray nuclear sources in low luminosity
elliptical nuclei are in M87 and NGC 6166 (Di Matteo et al. 2001). For
both of these nuclei, pure inflow ADAF models are consistent with the
observational constraints and can explain large fractions of the
observed nuclear fluxes.  {\it Chandra} observations of Sgr A$^{*}$
have also measured the X-ray flux of the nuclear point source
(Baganoff et al. 2001) and allowed us to estimate accretion rates onto
the central black hole (e.g. Narayan 2002; Quataert 2002). Sgr
A$^{*}$, like the elliptical nuclei, also requires low radiative
efficiency for the accreting gas and is consistent with ADAF
models. Further constraints on ADAF and inflow/outflow models in the
case of ellipticals will require knowledge of the contributions, over
all wavelengths, of emission from the bases of the jets.

\subsection{$\Mdot_{Bondi}$, overall energetics and activity}

Our fundamental conclusions rely on the assumption that the black hole
in M87 is accreting mass at the Bondi rate. The predicted nuclear
power of M87, for accretion at the Bondi rate, and $\eta=0.1$ is
$\sim$ 5 $\times$ $10^{44}
\ergps$ (see Eq.~4). This estimate roughly matches some 
estimates of the kinetic power in the jet which is
calculated to be $\approx 10^{44} \ergps$ (e.g.; Bicknell \& Begelman
1999; Owen et al. 2000; whereas Reynolds et al. 1996 obtain $\sim
10^{43} \ergps$).  A rough estimate of the jet power required to
evacuate, at the sound speed,
the inner cavities in the ICM associated with the
inner radio lobes ($\simeq 2-3 {\rm \kpc}$) in M87 implies 
$L_{\rm j} \sim 3 \times 10^{42} \ergps$ (Young et al. 2002).  
This value is smaller than the estimates of Bicknell
\& Begelman (1999) and Owen et al. (2000) but only considers
the cavities observed in the inner region of M87. Further, $L_{\rm j}$
would be higher if the cavities are evacuated supersonically.

We find, therefore, that the Bondi luminosity estimate roughly matches
the overall energetics of the black hole in M87 (if this is indeed
$\sim 10^{44} \ergps$). This may provide support for the relevance of Bondi
theory as a means of estimating the mass supply and the nuclear power
of black holes in ellipticals.  We note that the jet in M87 is also a
low-efficiency radiator with a total bolometric luminosity of the
order of $10^{42} \ergps$ (Owen et al. 2000).  Thus, based on these
estimates of the power of the nucleus, the energy input if all the
$\Mdot_{\rm Bondi}$ reaches the black hole, at present dominates
radiative losses from both the jet and accretion flow.

We discuss two alternative possibilities to the low-radiative
efficiency scenario: (a) as mentioned above, Blandford \& Begelman
(see also e.g. Stone et al. 1999) have suggested that low luminosity
accretion flows may develop strong outflows such that most of the mass
and energy is removed from the accretion flows.  In this case one may
speculate that the accretion energy is emerging in the jet (it has
been shown that the inner kiloparsec of M87 is highly magnetized and
turbulent, e.g. Owen, Eilek \& Keel 1990; Zhou 1998. Outflows may
become collimated and radiate more efficiently when interacting with
this medium). On the same note, Livio, Ogilvie \& Pringle (1999) have
argued that the energy extraction from black holes via MHD processes
such as the Blandford-Znajek mechanism (Blandford \& Znajek 1977) may
in-fact be most efficient for advection dominated flows (see also
Meier 2001).

(b) The central engine of M87 undergoes on-off activity cycles (e.g.,
Binney 1999; Owen et al. 2000). The black hole in M87 may have been
active for 100-200 Myr (as derived from estimates of the current age
of its radio halos; Owen et al. 2000). As long as the black hole is
fed (e.g. at the estimated Bondi rate) it will have a strong effect on
the ISM gas in the core (the bolometric X-ray luminosity of which is
$\sim 10^{43} \ergps$ in the inner 20 kpc or so; e.g. Nulsen \&
B\"ohringer 1995). The jet will disturb (or heat) the core regions,
support bulk flows and turbulence. These extra pressure forces may, in
turn, offset radiative cooling, support the gas against gravitational
infall and suppress accretion onto the black hole (hence the jet
activity itself). Once the central engine turns off, no energy is
deposited in the core so that accretion at the standard Bondi rate and
then jet activity resume.

The significant flattening of the X-ray gas density profile within the
central $3-4 \kpc$ (Fig.~2) may be a strong indication that indeed
energy is being deposited in the central region of M87 (see also
Loewestein et al. 2001 for other Virgo ellipticals). A flat density
profile can only arise if there is lots of mass dropout (see e.g.;
Quatert \& Narayan 2000) or maybe if gas is not flowing inward at
all. Given that mass dropout is not seen in X-ray data at larger radii
(B\"oehringer et al. 2002; Molendi \& Pizzolati 2001) it is possible
that something has affected the accretion radius and stifled
accretion.

It is interesting to note that the mild peak in the Figure 2 density
profile coincides with the ring at about 10 arcsec from the nucleus
(Figure 1) which is also where the H-$\alpha$ emission peaks. This
possibly indicates that there could be stifled inflow within this
radius and a cooling flow outside it. It is also notable that the few $
\kpc$ radius corresponds approximately to the size of the inner radio lobes
of M87 (Owen et al. 2000). There is also strong evidence of `cavities'
in the X-ray emitting gas of M87, the inner ones of which coincide with
the inner radio lobes (Young et al. 2002; see also Fig.~1) and a
number of other clusters containing radio galaxies (McNamara et
al. 2000; Fabian et al.  2000; Heinz et al. 2002); some of these
cavities are associated with observed radio lobes and others are
hypothesized to be relics of old radio activity.

It has been shown that Bondi accretion can be significantly suppressed
in a non-adiabatic gas (e.g. Ostriker et al. 1976). Following along
these lines, we show that, at present, the energy dumped into the ISM by
the jet in M87 may be sufficient to decrease the accretion rate with
respect to the Bondi value by a factor $\approxlt 10^6$. We do not
attempt to build a detailed model but just qualitatively discuss the
implications of non-adiabatic Bondi accretion for M87.

We consider a jet of power $L_{j}$ and velocity $v_{j}$ advancing into
the ISM with density $\rho$.  The jet forces its way outwards at a
speed $v_{h}$ which is obtained by balancing the outward momentum flux
of the jet with the pressure of the ISM. As long as $v_{h}$ is larger
than the sound speed $c_{s}$ in the ISM, the thermal pressure will be
dominated by the contribution $\rho v_{h}^2$ from ram pressure. Hence
$ v_{h}\sim ({L_{j}}/{\rho A_{h} v_{j}})^{1/2}$. When $v_{h}$ no
longer exceeds $c_{s}$, the jet will slow down and dump energy in the
surrounding medium. The inner radio lobes (in the inner few kpc) in
M87 may be associated with such a region.  Recent {\it Chandra}
observations of M87 (Young et al. 2002) and other radio galaxies in
clusters show no evidence that such interactions lead to sharp
bow-shock regions. More often, it seems that jets are responsible for
inflating bubbles in the surrounding ISM which then expand buoyantly
(i.e. roughly at the sound speed; e.g. Churazov et al. 2001; Reynolds,
Heinz \& Begelman 2002) and give rise to radio lobes and corresponding
X-ray structures.  We take $\pi r^2$ to be the cross sectional area of
the bubble inflated by the jet around the centre (which, for
simplicity, we take to be spherical see e.g. Churazov et al. 2001;
Reynolds et al. 2002)

Taking the mass, momentum and adding the energy conservation equation
 to the standard Bondi problem one can solve for the accretion radius
 (see details in Appendix A) with the addition of an energy (heat)
 source. In this case the accretion radius is reduced to:

\begin{equation} 
r_{\rm s} = \frac{GM}{2 c_{\rm s}^2} - 
\frac{2}{3}\frac{r_{\rm s}^2}{c_{\rm s}^3} (H),
\end{equation}
where the heating is in units of $\ergps \g^{-1}$ and the first term
in Equation 5 is the standard Bondi radius as in Equation 1.
According to the considerations above we write the energy deposited by
the jet as: $H= L_j/m$ (where m is the total gas mass) with $L_j \sim
\pi r^2 \rho v_j v_h^2$ so that $H \sim v_{j} v_{h}^2 / r$. In this
problem we take heating (i.e., the energy deposition) to be occurring
around the accretion radius of the black hole (see Appendix for
details). The inner radio lobes and X-ray cavities both on kpc scale
may be indications that this is not too bad an approximation (although
these are at radii which are more than an order of magnitude larger
than the accretion radius).  The resulting accretion radius will
rescale from the classical Bondi value roughly as,
\begin{equation}
r_{\rm s} \sim \frac{GM}{c_{\rm s}^2} \left(\frac{v_{j}}{c_{s}}\right)^{-1}
\end{equation}
where we have assumed that $v_{h} \sim c_{s}$, the velocity of the
bubble inflated by the jet in the ISM. Equation 6 clearly shows that
the if bulk motions (as well as heat) at speeds $> c_s$ are created by
the jet, then the accretion rate is affected.

The flow velocity of the M87 jet on scales of a few hundred parsec is
highly relativistic (Biretta, Zhou \& Owen 1995). Thus, according to
the above scaling the accretion radius of M87 is likely to be reduced
by a factor $\sim 1,000$. This would imply a decrease in mass
accretion rate by $\sim 10^6$. This is more than enough to relax the
requirement for low-radiative efficiency of the accreting gas. This
may also provide support for the idea that the engine of M87 may
undergo cycles of activity and at present the black hole is accreting
at rates much lower than the Bondi rate. However, the transit time
from the nucleus to the knot A in the jet (at a distance of $\sim 1$
kpc) is only about $\sim 3000$ yrs. This implies that for the
accretion rate to be low now, the last active phase must have been a
"burst-like" event of strong accretion on a relatively short
timescale.

\acknowledgements
We thank Eliot Quataert for comments on the manuscript.
T.\,D.\,M.\ acknowledges support for this work, while at CfA, provided
by NASA through Chandra Postdoctoral Fellowship grant number PF8-10005
awarded by the Chandra Science Center, which is operated by the
Smithsonian Astrophysical Observatory for NASA under contract
NAS8-39073 and for grant NAG-10105.  ACF and SWA aknowledge the Royal
Society.  This work was also supported by NASA through grants NAG
81755 and NAG 81027 to the University of Maryland.

\appendix
\section{Appendix}
In order to to show the effects of a jet of power $L_{\rm j}$ dumping
its energy in the ISM gas, we write the mass, momentum and energy
conservation equations for the accretion radius under the conditions
of spherically symmetric, steady ($\partial/\partial t =0$) accretion.
We consider the most of the heating to be occurring within or just outside the
accretion radius.

\begin{equation}
\Mdot = 4 \pi r^2 \rho(r) v(r),
\end{equation} 
\begin{equation}
v \frac{dv}{dr} + \frac{1}{\rho}\frac{dP}{dr} + \frac{GM}{r^2} = 0
\end{equation}
and
\begin{equation}
v\frac{d}{dr} \left[\frac{v^2}{2} + (\gamma - 1)\frac{P}{\rho}\right] =
-\frac{GM}{r^2}v + H
\end{equation}
where the quantities $\Mdot$, $\rho$, $v$, $P$ and $\gamma$ represent
the mass accretion, the mass density, the radial velocity, the
pressure, the gas adiabatic index, respectively.  $H$ is the heating
rate.  Using (A1) and taking $\gamma = 5/3$ we rewrite Eqs. (A2) and
(A3) as
\begin{eqnarray} 
\frac{dv}{dr} \left(\frac{v^2}{c_{\rm s}^2} -1 \right)& = &-\frac{GMv}{r^2c^2_{\rm s}} - \frac{2}{3} \frac{H}{c_{\rm s}^2} - \frac{2v}{r} \\ 
\frac{3}{2} \frac{dc_{\rm s}^2}{dr}\left(\frac{v^2}{c_{\rm s}^2} -1 \right)& = & \frac{GM}{r^2} - \frac{2 v^2}{r} + \frac{5v^2 -3c_{\rm s}^2}{3 c_{\rm s}^2 v} H \nonumber
\end{eqnarray}
For the flow to satisfy the appropriate boundary conditions, $v
\rightarrow 0$ as $r \rightarrow \infty$ and $ v
\rightarrow (GM/r)^{1/2}$ as $ r \rightarrow 0$, the two equations 
above must have a sonic point at $r = r_{\rm s}$ where $v^{2} =
c_{s}^2$. The only solutions that are continuous imply that the RHS of
both equations is 0. Both equation have the same sonic radius, and solving
for that gives,
\begin{equation} 
r_{\rm s} = \frac{GM}{2c_{\rm s}^2} - \frac{2}{3}\frac{r_{\rm s}^2}{c_{\rm s}^3} (H),
\end{equation}
where the heating and cooling rates are in units of $\ergps \g^{-1}$.
The factor $2/3$ applies to the $\gamma =5/3$ case but can be
substituted (and thus generalized) by the term $(\gamma -1)$.

\clearpage

\begin{figure}
\plotone{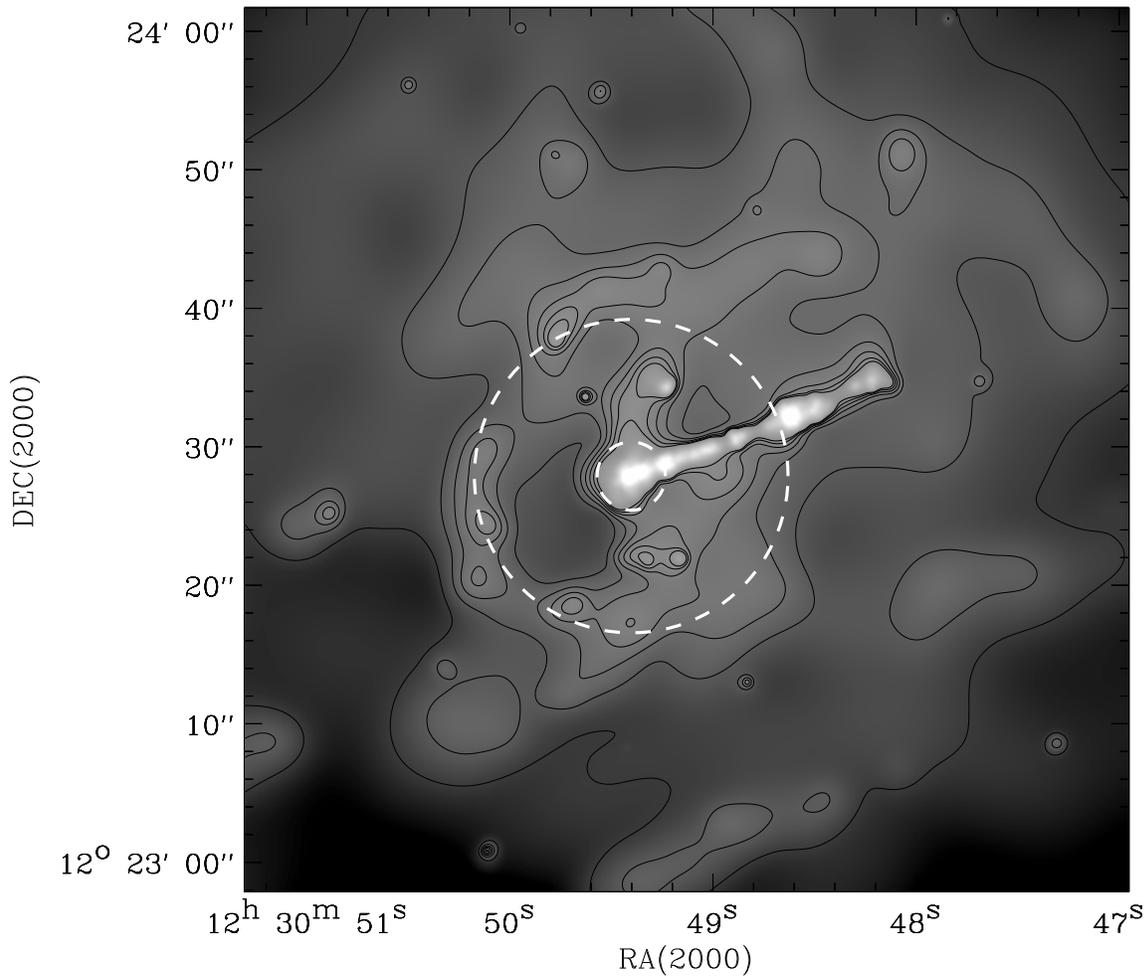}
\caption{\emph{Chandra} image of M87 and the core of the Virgo cluster in the
0.5 -- 5 keV energy band. Overlaid are 8 logarithmically spaced
contours from 5 -- 30 counts per $0\farcs5$ pixel (solid lines) and
the inner annulus used for the deprojection analysis (white dashed circles,
centered on the nucleus). The image has been adaptively smoothed so
that the signal-to-noise ratio of the signal under the smoothing
kernel at each point on the image is at least 3.}
\end{figure}
\clearpage

\begin{figure}
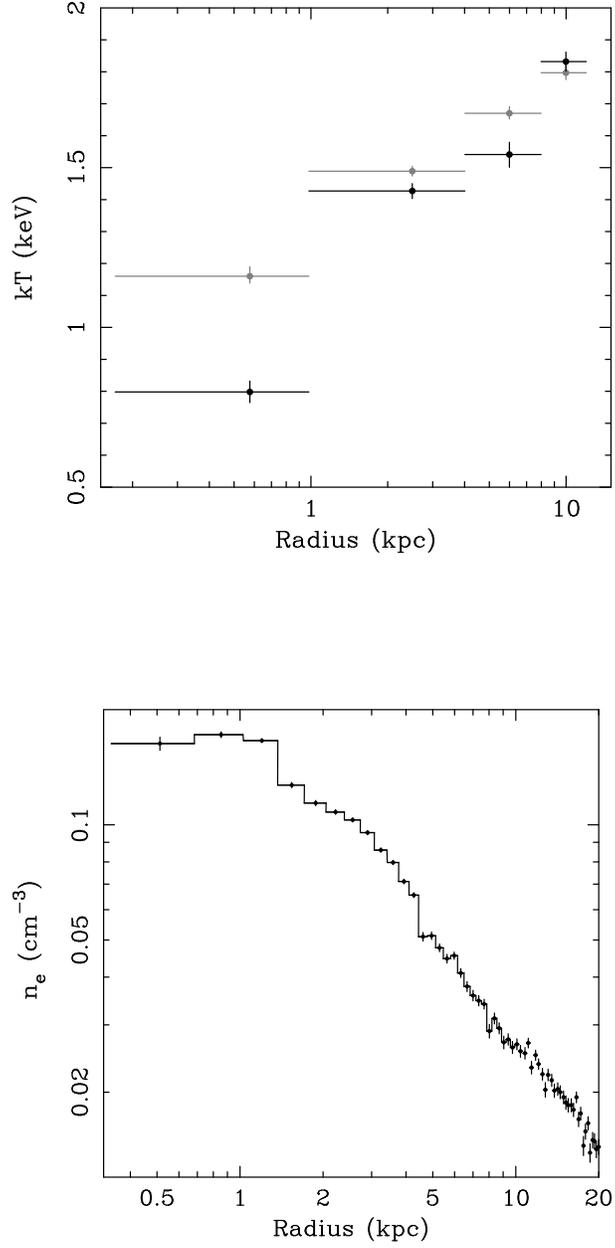

\centerline{\vbox{ 
\psfig{file=f2a.eps,width=8.0truecm,angle=270}
\vspace{2cm}
\psfig{file=f2b.eps,width=8.0truecm,angle=270}}}
\caption
{Top: The temperature profile of the ISM in M87 measured by {\it
Chandra}. The dark points show the deprojected temperature
measurements and the grey ones the observed (projected) values. The
$1\sigma$ error bars have been multiplied by a factor of 3 to improve
their visibility. Bottom: The (electron) density profile of the
ISM. Error bars are the $1\sigma$ uncertainties.}
\end{figure}

\clearpage

\begin{figure}
\psfig{file=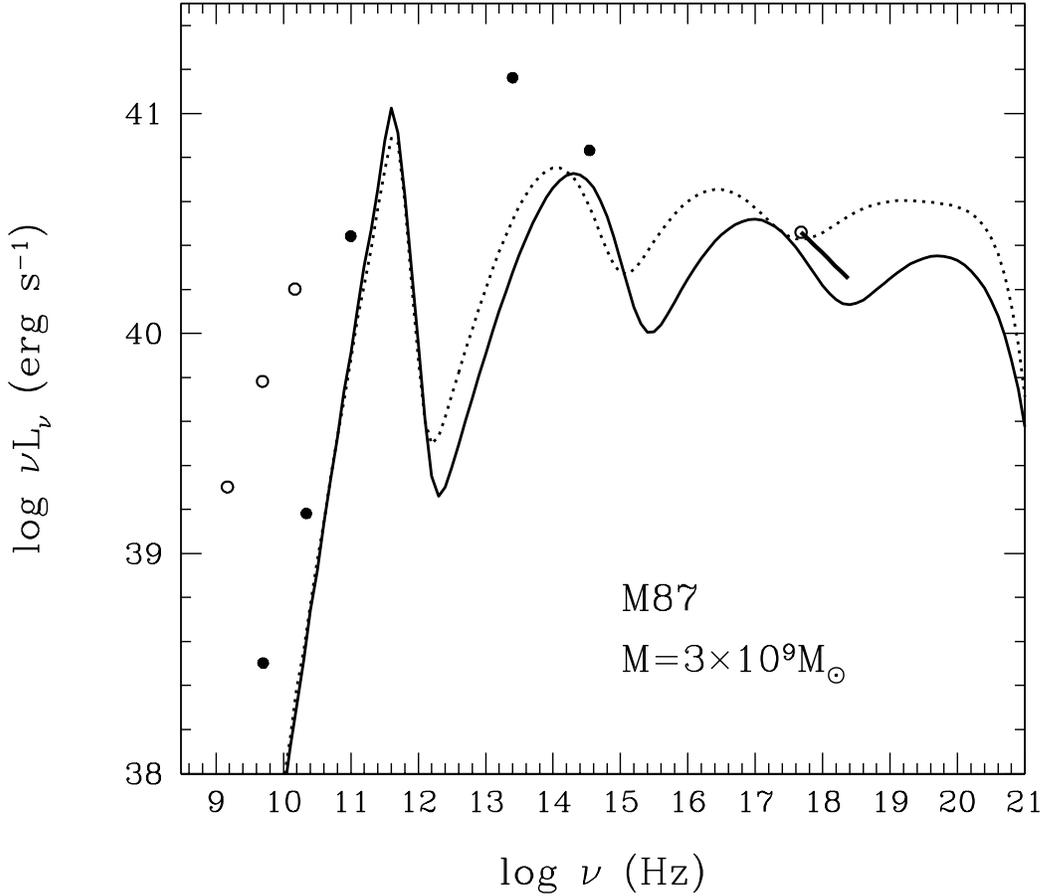,width=16.0truecm}
\caption{Spectral models calculated for hot accretion flows normalized to the
{\it Chandra} X-ray flux. The solid line is an ADAF model with
$\delta=0.3$. The required accretion rate, $\dot{m} = 7 \times
10^{-4}$, is consistent with the Bondi value. The dotted line shows a
model with $\delta = 0.01$ with $\dot{m}= 10^{-3}$. The filled dots
are the VLBI nuclear flux density measurements by Paulini-Toth et
al. (1981); Spencer \& Junor (1986) and B{\"a}{\"a}th et al. (1992) in
increasing frequency order, respectively. The IR, 10 $\mu$m nuclear
flux measurement is the best IR nuclear flux limit (resolution of
$\sim 0.46$ arcsec) from Gemini (Perlman et al. 2001). The optical
nuclear continuum measurement (also a filled dot) is from HST (which
has resolution of 0.15 arcsec) observations by Harms et
al. (1994). The open circles are lower resolution measurements by
Biretta et al. (1991). The radio and optical data are tabulated by
Reynolds et al. (1996). The thick solid line shows the Chandra spectrum (in
the $0.2-10$ keV~band).}
\end{figure}

\end{document}